\pdfoutput=1
\documentclass[preprint,12t,1p,times,fleqn]{elsarticle}  

\journal{Int. J. Numer. Methods Fluids}

\usepackage{hyperref}
\hypersetup{%
  colorlinks=true,
  pdfborder={0 0 0 },
}

\usepackage[list-units=single,separate-uncertainty=true,multi-part-units=single]{siunitx}
\usepackage{amsmath}
\usepackage{amssymb}
\usepackage{bm}
\usepackage{booktabs}

\usepackage[capitalize,nameinlink]{cleveref}
\crefdefaultlabelformat{#2\textbf{#1}#3} 

\crefname{figure}{\textbf{Fig.}}{\textbf{Figs.}}
\Crefname{figure}{\textbf{Figure}}{\textbf{Figures}}
\crefname{table}{\textbf{Table}}{\textbf{Tables}}
\Crefname{table}{\textbf{Table}}{\textbf{Tables}}
\crefname{section}{\textbf{Sec.}}{\textbf{Sections}}
\Crefname{section}{\textbf{Sec.}}{\textbf{Sections}}
\crefname{subsection}{\textbf{Sec.}}{\textbf{Sections}}
\Crefname{subsection}{\textbf{Sec.}}{\textbf{Sections}}

\usepackage{listings,newtxtt}

\usepackage{caption}
\usepackage{multirow}
\usepackage{tabularx}
\newcolumntype{C}{>{\centering\arraybackslash}X}     
\newcolumntype{R}{>{\raggedleft\arraybackslash}X}    
\newcolumntype{L}{>{\raggedright\arraybackslash}X}   

\newcommand{\pdif}[2]{\frac{\partial #1}{\partial #2}}
\usepackage{xspace}
\newcommand{\etal}{\textit{et~al}.\xspace}
\newcommand{\bol}[1]{\boldsymbol{#1}}
\newcommand{\order}[1]{\mathcal{O}(#1)}

\begin{document}

\begin{frontmatter}

\title{Physics-informed neural network applied to  \\ \relax  surface-tension-driven liquid film flows}

\author[M]{Yo Nakamura\corref{}}
\author[B]{Suguru Shiratori\corref{cor1}}
\ead{sshrator@tcu.ac.jp}
\author[B]{Ryota Takagi\corref{}}
\author[B]{Michihiro Sutoh}
\author[B]{Iori Sugihara}
\author[B]{Hideaki Nagano}
\author[B]{Kenjiro Shimano}

\cortext[cor1]{Corresponding author}

\affiliation[M]{%
organization={Graduate School of Integrative Science and Engineering, Tokyo City University},
city={Setagaya-ku},
state={Tokyo},
postcode={158-8557},
country={Japan}
}

\affiliation[B]{%
organization={Department of Mechanical Systems Engineering, Tokyo City University},
city={Setagaya-ku},
state={Tokyo},
postcode={158-8557},
country={Japan}
}

\begin{abstract}
A physics-informed neural network (PINN), which has been recently proposed by Raissi \etal\,  [{\it J. Comp. Phys.} 378, pp. 686-707 (2019)], 
is applied to the partial differential equation (PDE) of liquid film flows.
The PDE considered is the time evolution of the thickness distribution $h(x,t)$ owing to the Laplace pressure,
which involves 4th-order spatial derivative and 4th-order nonlinear term.
Even for such a PDE, it is confirmed that the PINN can predict the solutions with sufficient accuracy.
Nevertheless, some improvements are needed in training convergence and accuracy of the solutions.
The precision of floating-point numbers is a critical issue for the present PDE.
When the calculation is executed with a single precision floating-point number, 
the optimization is terminated due to the loss of significant digits.
Calculation of the automatic differentiation (AD)
dominates the computational time required for training and becomes exponentially longer with increasing order of derivatives.
By splitting the original 4th-order single PDE into lower-order coupled PDEs,
the computational time for each training iteration is greatly reduced.
The sampling density of training data also significantly affects training convergence.
For the problem considered in this study, 
improved convergence was obtained
by allowing the sampling density of training data to be greater in earlier time ranges, 
where the rapid flattening of the thickness occurs.
\end{abstract}


\end{frontmatter}

\section{Introduction}
Thin liquid films are ubiquitous in nature and technology \cite{Oron1997,Craster2009}.
The coating process of photoresist films plays important roles in industrial micro-fabrication techniques.
For semiconductor devices, coated photoresist films are removed in the middle of the fabrication process,
whereas for MEMS devices, anti-reflection films for optical lenses, or color filters of image sensors,
some parts of photoresist films are used as final structures of the devices.
Therefore, the design of devices determines the film thickness,
which often exceeds the order of \SI{10}{\micro\meter}, and its uniformity is desired at a high level.
However, liquid film suffers from various thickness undulations
due to physical phenomena during the coating process.
The better-known thickness undulations are ``edge-bead'' and ``striation''.
Edge-bead is a thick ridge that appears along the substrate periphery.
The substrate region eroded by edge beads is not usable, thus they result in the loss of product yield.
Shiratori and Kubokawa investigated edge-bead generation for the case
where the bead had a double-peaked shape in the direction away from the substrate periphery,
and proposed a simple explanation for the mechanism of the double peak \cite{swan-POF-2015}.
Striation is another typical thickness undulation, 
which appears as radial spoke-like patterns all over the film \cite{Birnie2001,Haas-PhD-2006}.
The mechanism of striation has been investigated and, 
Marangoni-B\'enard instability is considered to play a fundamental role 
\cite{Birnie2001,Haas-PhD-2006,Daniels1986,Birnie1997,Haas2000,Haas2002,Kozuka2004,Birnie2005,Birnie2013}.
Shiratori \etal\, investigated the formation process of striations 
by measuring spatio-temporal thickness variations 
during a spin coating for mixtures of epoxy resin and xylene solvent \cite{swan-HMT-2020}.
They found that short-wavelength thickness variation may suddenly vanish in the middle of the process.
They also performed a numerical simulation to predict the transient Marangoni number during spin-coating,
and concluded that the sudden change in the thickness spectrum found in the experiment
is associated with the decrease in the Marangoni number to a sub-critical value for a specific bifurcation of the flow regimes.

Such thickness variations must be avoided or suppressed in industry.
To this end, numerical simulations are often used to find the optimal coating conditions, 
which allows thickness undulations to be minimized.
In previous research on thin liquid films,
many numerical methods have been developed \cite{%
Weidner1996,%
Diez2002,%
Roy2002,%
Schwartz2004,%
Sultan2005,%
Gaskell2006,%
Yiantsios2006%
}.
However, it has been difficult to solve the inverse problem of finding the optimal coating conditions, because of the computational cost of the time integration of the governing equations.
In the general optimization procedure, 
it is necessary to evaluate the objective function and its gradient with respect to the parameters to be optimized.
Both require considerable computational cost in conventional numerical methods.

A fast-growing machine learning approach can be considered to resolve such problems.
Previously,
machine learning approaches have been applied to the field of fluid dynamics
\cite{%
Ling2016,%
Brenner2019,%
Raissi2020,%
Fukami2021,%
Nakamura2021%
}.
There are many potential applications including
not only a surrogate model for high-fidelity simulation, 
but also the reduced-order modeling, closure modeling, or a flow control.
The recent trend of the machine-learned fluid dynamics are summarized by, e.g. Brunton \etal \cite{Brunton_2020},
Fukami \etal \cite{Fukami2020} and Duraisamy \cite{Duraisamy2021}.
In general, machine learning has been successful in fields such as image recognition and natural language processing, where it is easy to obtain a sufficiently large training dataset.
The above-mentioned applications of machine learning to fluid dynamics primarily involve turbulence, 
for which abundant computational results have been accumulated.
 In fields where sufficient training data is not readily available, obtaining training data for numerical simulations is typically very expensive. 
This applies to research into liquid film flows.

In this study,
we focus on a physics-informed neural network (PINN), as recently proposed by Raissi \etal\cite{Raissi_JCP_2019}.
The PINN learns the solutions of a partial differential equation (PDE) for a given dataset.
In the training process for a PINN, a loss function is defined as the mean square error of the predicted solutions of the PDE.
To evaluate the loss function, the temporal and spatial derivatives of the unknowns are calculated 
by automatic differentiation (AD), which is implemented in the neural network (NN) framework. 
Once a PINN has been trained, the solutions for any time instance 
can be calculated directly without time integration by forward computation by the NN.
In addition, the gradient of the solution with respect to the input variable can be calculated efficiently using AD.
The methodologies of the PINN are based on supervised learning; nevertheless, it does not require supervisor data,
because the supervisor is assigned to the governing equation which must equate to zero.
Therefore, the PINN is expected to perform well with limited training data.

However, the application and validation of a PINN has only been carried out for several simple problems.
In particular, there is no prior research regarding applications of a PINN to liquid film flows.
The typical governing equation for liquid film flows involves
a fourth-order spatial derivative and fourth-order nonlinearity in the Laplace pressure term.
For higher-order derivatives, the AD may require greater computational time.
The Laplace pressure term behaves as a diffusion of the curvature with the cube of the thickness $h^3$ being the diffusion coefficient.
Thus, if $h$ is unexpectedly predicted as negative, 
the Laplace pressure term acts as an anti-diffusion, which leads to numerical instability.
For equations in which this occurs, the applicability of a PINN has not been investigated with respect to accuracy and computational cost.
The goal of the present study is to resolve this problem.
We aim to clarify 
whether a PINN can predict the solution of the governing equation with sufficient accuracy and acceptable computational time.
For validation, the solution predicted by the PINN is compared with that calculated by the conventional finite difference method (FDM).
We also investigate the computational cost for the automatic differentiation of high-order derivatives.
If excessive computational time is required, 
it may be preferable to divide the equation into multiple sub-equations of low-order derivatives, by introducing intermediate unknowns.

After the problem formulation given in \cref{sec-formulation},
the detailed architecture and numerical methods are formulated in \cref{sec-PINN}.
Detailed methods in the FDM, which is executed for comparisons, are described in \cref{sec-FDM}.
The validity and efficiency of the PINN for liquid film flows are confirmed in \cref{sec-result}
by comparing the results obtained in this study with those obtained using the FDM. 
In \cref{sec-result}, in addition to the representative case, 
several aspects are investigated to improve the performance of the PINN.

\section{Problem formulation}\label{sec-formulation}
\subsection{Basic form}
\begin{figure}[tbp]
\centering\includegraphics[width=0.7\textwidth]{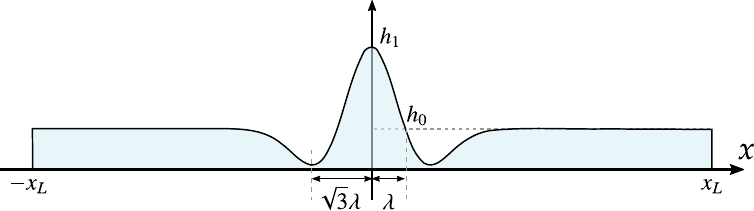}
\caption{\label{fig-problemSetup}
System to be considered.
A liquid film of average thickness $h_0$ is initially bumped with maximum height $h_1$,
and decays due to Laplace pressure.
}
\end{figure}

We consider a liquid film of an incompressible Newtonian fluid of viscosity $\mu$ and surface tension $\sigma$,
which is coated on the substrate with average thickness $h_0$.
As shown in \cref{fig-problemSetup},
the film is initially bumped according to the following function:
\begin{equation}
h_\text{ini}(x) = \left(h_1 - h_0\right) \left( 1 - \left( \frac{x}{\lambda} \right)^2 \right) \exp\left(-\frac{x^2}{2\lambda^2}\right) + h_0, \label{eq-Ini}
\end{equation}
where $h_1$ is the maximum thickness,
and $\lambda$ is the parameter which determines the extent of the bump.
Such a deformed interface has a curvature distribution, which causes a gradient in the Laplace pressure.
Owing to the flow driven by the pressure gradient, the interface deformation decays with time.

The problem addressed in this study may seem rather simple.
In the practical liquid film flows, 
there may occur some physics other than the Laplace pressure, 
such as the advection and leveling effects of the gravity,
the Marangoni convection, 
a wetting phenomena, or evaporation and so on.
These physics can be expressed by adding convection or diffusion terms in the governing equation of the present work.
In addition, the framework of the PINN can be applied to two or three dimensional problems.
Nevertheless, the present study focuses only on the Laplace pressure in one-dimensional problems. 
In the previous studies, it has been confirmed that
the PINN is able to predict accurate solutions of several fundamental types of transport equations,
for instance, Burgers equation, Navier-Stokes equation and Allen-Cahn equation.
These equations involve the advection and diffusion terms 
which are described by the spatial derivatives up to the second-order,
whereas the Laplace pressure term in the liquid film flow is described by the fourth-order derivative.
The main purpose of the present study is to confirm 
whether the PINN can be applied to a such a higher-order PDE, 
and to extract the computational issues to which the attention should be paid.

The leveling process of the thickness undulation can be modeled using the following governing equation
\begin{equation}
F: \pdif{h}{t} + \pdif{}{x} \left( h^3 \frac{\partial^3 h}{\partial x^3} \right) = 0,  \label{eq-governing}
\end{equation}
where $h$ is the thickness of the liquid film, 
$t$ is the time, and $x$ is the spatial coordinate.
The derivation of \cref{eq-governing} is described, for instance, in Oron \etal \cite{Oron1997} or Shiratori \etal \cite{swan-HMT-2020}.
The equations are nondimensionalized using the scale shown in \cref{tab-scales}.
The spatial domain is defined as $x \in [-x_L:x_L]$ 
and the following boundary conditions are employed:
\begin{equation}
\pdif{h}{x} = \frac{\partial^3 h}{\partial x^3} = 0  \qquad \text{at} \quad  x = \pm x_L,   \label{eq-BC1}
\end{equation}
which are corresponding to symmetry conditions for the thickness $h$ and the curvature $\partial_x^2 h$, respectively.

\begin{table}[bp]
\caption{\label{tab-scales}Scales of the dimensionless variables.}
\centering
\begin{tabular}{cccc}
\toprule
Lateral length  & Thickness   & Time                        & Pressure               \\
\midrule
 $\lambda$      & $h_0$       & $3\mu \lambda^4 / (\sigma h_0^3)$ & $h_0 \sigma / \lambda^2$     \\
\bottomrule
\end{tabular}
\end{table}

\subsection{Division of equations}
\Cref{eq-governing} involves the fourth-order derivative, which may incur a large computational cost in the automatic differentiation step in the PINN.
To address this, instead of using the original \cref{eq-governing}, we divide the equation by introducing intermediate variables.
The second-order two PDEs can be obtained by explicitly introducing the Laplace pressure $p$ as
\begin{subequations}
\label{eq-governingEq2}
\begin{align}
G_1:& \pdif{h}{t} - \pdif{}{x} \left( h^3 \frac{\partial p}{\partial x} \right) = 0,  \\
G_2:& p           + \frac{\partial^2 h}{\partial x^2} = 0.
\end{align}
\end{subequations}
Similarly, the original PDE can be divided into first-order four PDEs as
\begin{subequations}
\label{eq-governingEq4}
\begin{align}
H_1:& \pdif{h}{t} - 3h^2 q_1 q_3 - h^3 \pdif{q_3}{x} = 0,  \\
H_2:& q_1 - \pdif{h}{x} = 0, \\
H_3:& q_2 - \pdif{q_1}{x} = 0, \\
H_4:& q_3 - \pdif{q_2}{x} = 0,
\end{align}
\end{subequations}
by introducing additional variables $q_1$, $q_2$, and $q_3$.
Regarding the initial condition for the divided equations, 
\cref{eq-Ini} is applied to $h$, whereas no conditions are applied to the intermediate variables $p$, $q_1$, $q_2$, and $q_3$.
In the boundary conditions,
the curvature gradient condition $\partial_x^3 h = 0$ is replaced by the lower-order derivative forms:
\begin{align}
\pdif{h}{x} = \pdif{p}{x} = 0 & \qquad \text{at} \quad  x = \pm x_L,   \qquad \text{(for 2nd-order problem)}   \label{eq-BC2} \\
\pdif{h}{x} = q_3 = 0         & \qquad \text{at} \quad  x = \pm x_L,   \qquad \text{(for 1st-order problem)}   \label{eq-BC4} 
\end{align}

\section{Physics-informed neural networks}\label{sec-PINN}

\subsection{Architectures}
\begin{figure}[tbp]
\centering\includegraphics[width=\textwidth]{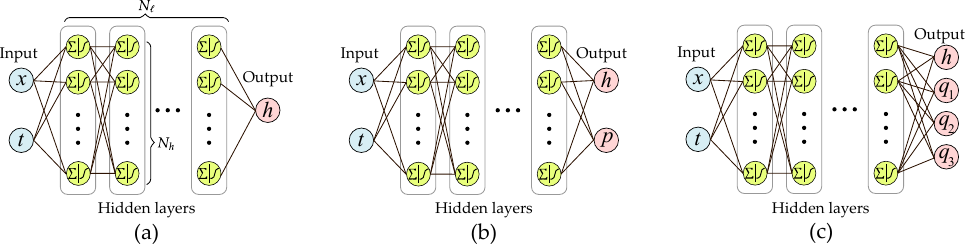}
\caption{\label{fig-structure}
Network structures of the PINNs used in the present study.
(a) single-output network for the 4th-order equation, 
(b) two-output network for the 2nd-order equations,
and
(c) four-output network for the 1st-order equations.
}
\end{figure}

Throughout this study,
we have been using simple deep feed-forward neural network architectures, as shown in \cref{fig-structure}.
All the hidden layers are fully connected dense layers
and all the activation functions are hyperbolic tangents \texttt{tanh}.
For the fourth-order governing equation \cref{eq-governing},
a single-output network (\cref{fig-structure}(a)) is used,
whereas a two-output network (\cref{fig-structure}(b)) and
a four-output network (\cref{fig-structure}(c))
are used for the second-order PDEs \cref{eq-governingEq2} and first-order PDEs \cref{eq-governingEq4}, respectively.
The number of hidden layers is defined as $N_\ell$
and the number of neurons in each hidden layer is denoted as $N_h$.
Thus, the total number of the training parameters is
$(N_\text{in} + 1) N_h + (N_\ell - 1)(N_h + 1)N_h + (N_h + 1) N_\text{out}$,
where $N_\text{in}$ and $N_\text{out}$ are the number of inputs and outputs, respectively.

\subsection{Loss function}
The loss function for training is defined as follows:
\begin{subequations}
\begin{align}
 J    &= J_f  +  J_b + J_o  +  J_p,   \\
 J_f  &= \frac{1}{N_f} \sum_{i=1}^{N_f}  \sum_{k=1}^{N_k} \left[\mathcal{G}_k\left(t_f^i, x_f^i \right)\right]^2 ,   \label{eq-Jf}  \\
 J_b  &= \frac{1}{N_b} \sum_{i=1}^{N_b}\left( \left[ \mathcal{B}\left(t_b^i, x_b^i \right) -b_1^i \right]^2 
                                          +\left[ \partial_x   h\left(t_b^i, x_b^i \right) -b_2^i   \right]^2 \right),  \\
 J_o  &= \frac{1}{N_o} \sum_{i=1}^{N_o} \left[ h\left(t_o^i, x_o^i \right) - h_o^i \right]^2,  \\
 J_p  &= \frac{1}{N_f} \sum_{i=1}^{N_f} f_p\left(h\left(t_f^i, x_f^i \right)\right),
\end{align}
\end{subequations}
where $J_f$ is the mean squared error (MSE) of the governing equations $\mathcal{G}_k(t,x)$.
Although the governing equations must be mathematically satisfied,
residuals remain when the equation is evaluated numerically by a neural network using automatic differentiation.
The components $J_b$ and $J_o$ are the MSEs of the boundary and initial conditions, respectively.
$\{t_f^i, x_f^i\}_{i=1}^{N_f}$ denote the collocation points for the governing equation,
whereas $\{t_b^i, x_b^i, b_1^i, b_2^i\}_{i=1}^{N_b}$
and $\{t_o^i, x_o^i, h_o^i\}_{i=1}^{N_o}$ are the boundary and initial conditions of the training data set, respectively.
The values for the boundary and initial conditions are described in the next subsection.

The concrete forms of $\mathcal{G}_k$ and $\mathcal{B}$ are dependent on the problem, as shown in \cref{tab-problems}.
Here, the number of PDEs is denoted by $N_k$.
\begin{table}[bp]
\caption{\label{tab-problems} Correspondence of equations depending on the problem.}
\centering
\begin{tabular}{cccc}
\toprule
                     & Governing PDEs         & Number of PDEs     & Curvature boundary condition    \\
Problem              & $\mathcal{G}_k(t,x)$   &  $N_k$             & $\mathcal{B}(t,x)$    \\
\midrule
4th-order one  PDE   & $F(t,x)$               & \num{1}            & $\partial^3_x h$ \\
2nd order two  PDEs  & $G_k(t,x)$             & \num{2}            & $\partial_x p$ \\
1st-order four PDEs  & $H_k(t,x)$             & \num{4}            & $q_3$ \\
\bottomrule
\end{tabular}
\end{table}
The evaluation of $J_b$ and $J_o$ requires supervisor data,
whereas the term $J_f$ does not need supervisor.
$J_p$ is the loss based on a penalty function, defined as
\begin{equation}
 f_p(h) = \max \left( 0, \exp\left(- h\right) -1 \right), \label{eq-penalty}
\end{equation}
which is introduced so that the network does not predict a negative value of $h$.
In the governing equations \cref{eq-governing,eq-governingEq2,eq-governingEq4}
the fourth-order terms represent the diffusion of the Laplace pressure
with the term $h^3$ being the diffusion coefficient.
Thus, if the PINN unexpectedly predicts a negative value of $h$,
the equation behaves as anti-diffusion, rendering computation unstable.
The penalty function \cref{eq-penalty} is introduced in the loss function to avoid this situation.

\subsection{Training data}\label{sec-trainingData}
\begin{figure}[tbp]
\centering\includegraphics[width=0.6\textwidth]{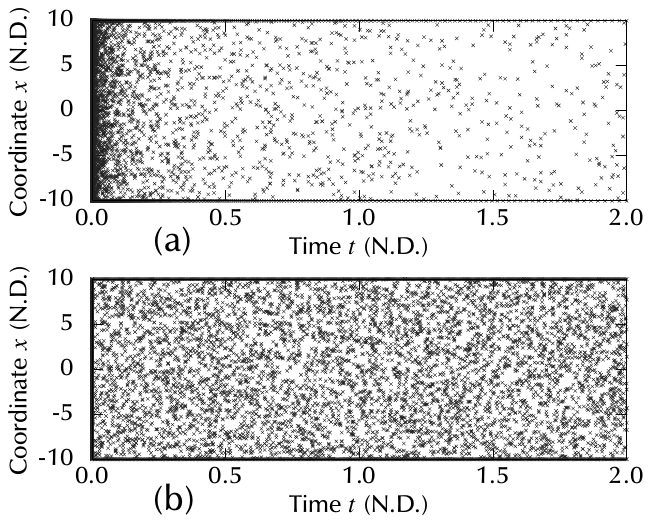}
\caption{\label{fig-samplePoints}
Distributions of collocation points $(t_f, x_f)$ for the evaluation of governing equations
(a) logarithmically sampled time points, and
(b) uniformly sampled time points.
The data sets $(t_o, x_o)$ and $(t_b, x_b)$ are also plotted.
}
\end{figure}
The collocation points $\{t_f^i, x_f^i\}_{i=1}^{N_f}$ for the governing equation
are generated according to the following procedures.
First,
one-dimensional arrays are generated
for both time $\{t_s\}_{s=1}^{N_t}$ and space $\{x_s\}_{s=1}^{N_x}$.
The array $x_s$ is generated with equidistant space as
\begin{equation}
x_s = x_L \left(-1 + 2\left(\frac{s - 1}{N_s - 1}\right)\right), \qquad s = 1, \cdots, N_s,
\end{equation}
whereas the array for the time points is evenly spaced on a logarithmic scale as
\begin{equation}
\left.
\begin{split}
a_s  &= \log_{\Gamma}\left(\frac{t_\text{min}}{t_\text{max}}\right) \left(\frac{N_s - s}{N_s - 1} \right) \\
t_s  &= t_\text{max} \, \Gamma^{a_s}
\end{split}
\quad \right\}
 \qquad s = 1, \cdots, N_s, \label{eq-logScale}
\end{equation}
where
$\Gamma$ is the base of the logarithm,
$t_\text{min}$ is the smallest time interval, and
$t_\text{max}$ is the upper bound of the time range under consideration.
For comparison, uniform time spacing
\begin{equation}
t_s = t_\text{max} \left(\frac{s - 1}{N_s - 1}\right), \quad s = 1, \cdots, N_s,  \label{eq-linScale}
\end{equation}
is also investigated.
Using these one-dimensional arrays $\{t_s\}_{s=1}^{N_t}$ and $\{x_s\}_{s=1}^{N_x}$,
two-dimensional $N_t \times N_x$ grid points are generated by combining two arrays in round-robin style.
From these parent grid points, the final collocation points of $N_f$ are randomly sampled.
The size of the parent arrays is $N_t \times N_x$, which is sufficiently larger than the final collocation points $N_f$.
Example collocation points are shown in \cref{fig-samplePoints}.

The training dataset for the boundary condition $\{t_b^i, x_b^i, h_b^i\}_{i=1}^{N_b}$
is generated as follows:
\begin{equation}
\left.
\begin{split}
a_i  &= \log_{\Gamma}\left(\frac{t_\text{min}}{t_\text{max}}\right) \left(\frac{N_b - i}{N_b - 1} \right) \\
t_b^i &= t_\text{max}\, \Gamma^{a_i} \\
x_b^i &= \pm x_L \\
b_1^i &= b_2^i = 0
\end{split}
\,\, \right\}
\quad i = 1, \dots, N_b.
\end{equation}
The same condition is applied at both boundaries $x = \pm x_L$;
Thus, the total size of the data set for the boundary conditions is $2 N_b$.
For the initial condition, the training dataset $\{t_o^i, x_o^i, h_o^i\}_{i=1}^{N_o}$
is generated on the equally-spaced spatial grids as:
\begin{equation}
\left.
\begin{split}
t_o^i &= 0  \\
x_o^i &= x_L \left( -1 + 2\left(\frac{i-1}{N_o-1}\right)\right) \\
h_i^i &= h_\text{ini}(x_o^i) 
\end{split}
\,\, \right\}
\quad i = 1, \dots, N_o,
\end{equation}
where $h_\text{ini}(x)$ is the function defined in \cref{eq-Ini}.

\subsection{Implementation}
For optimization we selected the L-BFGS-B method, which is a quasi-Newton,
full-batch gradient-based optimization algorithm \cite{Liu1989}.
The code for the PINN is implemented in \texttt{Python} 
using the \texttt{tensorflow-1.8} machine learning library \cite{tensorflow2016},
and executed with assistance of GPUs of the NVIDIA Tesla P100 and the NVIDIA A100.

\section{Finite difference method}\label{sec-FDM}
To confirm whether the PINN can correctly predict the solution, 
the same problem is also calculated using the finite difference method (FDM).
The governing equation \cref{eq-governing} is discretized by finite differences 
on an equidistant grid of $N_x$ points with an interval $\Delta x$.
The positivity-preserving scheme\cite{Zhornitskaya1999} is applied to 
$h^3$ of the Laplace pressure term in \cref{eq-governing},
because the standard finite difference schemes do not necessarily preserve
positivity of $h$, and the occurrence of non-positive solutions introduces
artificial instability \cite{Diez2000}.
Due to the fourth spatial derivative of $h$ in \cref{eq-governing},
a time step restriction for the numerical stability requires
$\Delta t < \mathcal{O}\left( \Delta x^4 \right)$, which is a severe condition in practice.
To cope with this restriction,
the fully implicit time integration of the Newton-Kantrovich method is applied
with the Crank-Nicholson method, according to Diez and Kondic \cite{Diez2002}.
The resulting linear systems are solved using BiCGStab iterations \cite{VanDerVorst:BiCGStab:1992} 
with an auto-accelerated incomplete LU factorization (ILU) preconditioner \cite{Miki2013}.
The code for the FDM is implemented in \texttt{C++} with OpenMP parallelization and 
executed using an Intel Xeon E5-2695v4 CPU.
The accuracy of the present FDM solver has been
validated through comparison with the benchmark problems provided by Dies and Kondic\cite{Diez2002}.
Although the result of the FDM is not an exact analytical solution, 
the prediction by the PINN is validated by comparing it with the FDM's result, 
which is calculated with sufficiently high resolution for the spatial grid ($\Delta x = \num{5e-3}$) and time step ($\Delta t = \num{4e-5}$).

\section{Results and discussions}\label{sec-result}
\subsection{Representative case}\label{sec-repCase}
\begin{table}[tbp]
\caption{\label{tab-setup1}
Methods employed.}
\centering
\begin{tabular}{ll}
\toprule
Setting                       &  Base value  \\
\midrule
Activation function                  &  $\tanh$ \\
Precision of floating-point numbers  &  FP64  \\
Type of collocation time points      &  log-scale \\
Method of optimization               &  L-BFGS-B \\
\bottomrule
\end{tabular}
\end{table}
\begin{table}[tbp]
\caption{\label{tab-setup2}
Parameters and representative values used.}
\centering
\begin{tabular}{lcc}
\toprule
Parameter     & Symbol                      &   Value \\
\midrule
Time range                           & $t_\text{max}$  &   \num{2}    \\
Spatial extent                       & $x_L$           &   \num{10}   \\
Maximum height                       & $h_1 / h_0$     &   \num{3}    \\
Data size for bulk equation          & $N_f$           &   \num{6000} \\
Data size for boundary conditions    & $N_b$           &   \num{1000} \\
Data size for initial conditions     & $N_o$           &   \num{500} \\
Number of hidden layers              & $N_\ell$        &   \num{8} \\
Number of neurons in each layer      & $N_h$           &   \num{20} \\
\bottomrule
\end{tabular}
\end{table}

\begin{figure}[tbp]
\centering\includegraphics[width=0.7\textwidth]{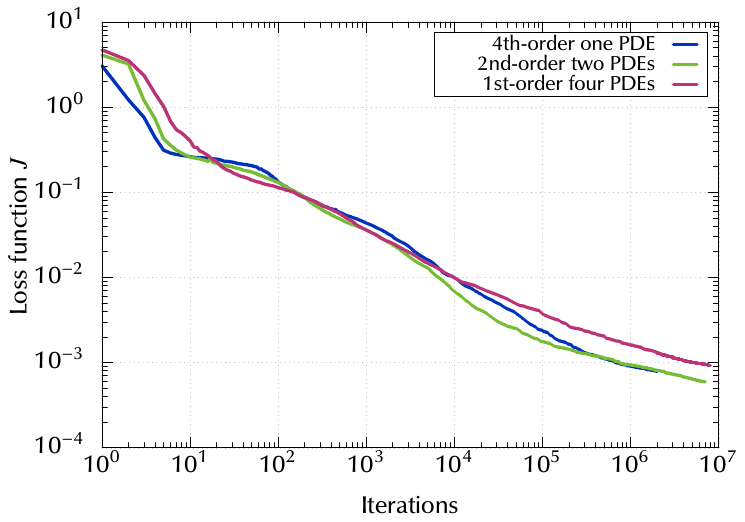}
\caption{\label{fig-lossEq124}
History of the loss function during training.
The blue line indicates the result for the representative case shown in \cref{tab-setup1,tab-setup2}.
The green and red lines depict the cases where the 
governing equation is divided into 
two 2nd-order PDEs and
four 1st-order PDEs, respectively.
}
\end{figure}

\begin{figure*}[tp]
  \centering\includegraphics[width=\textwidth]{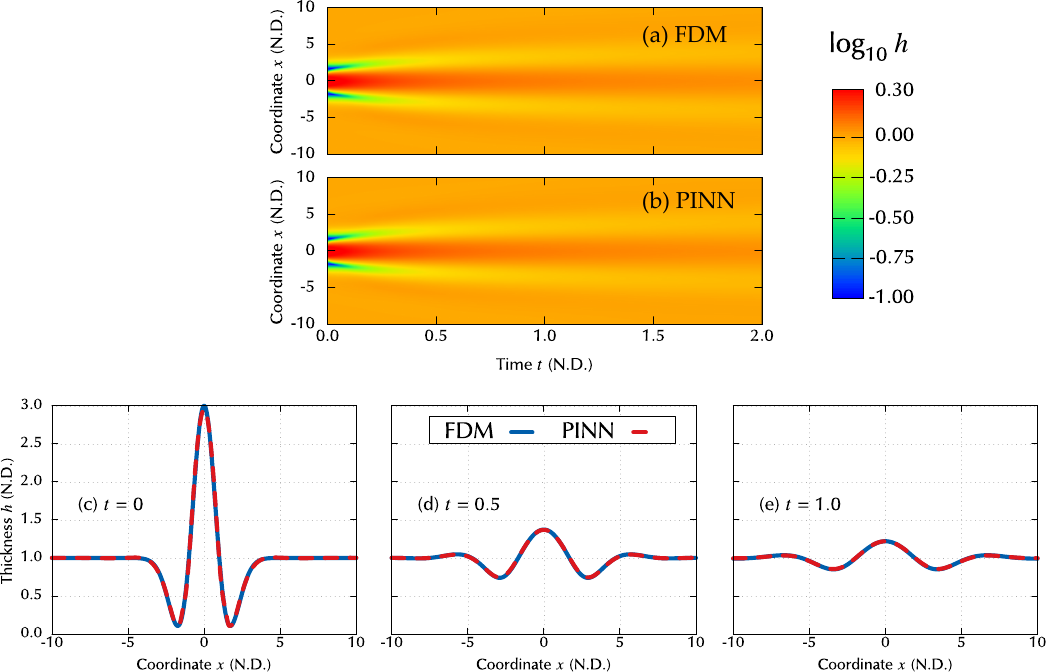}
\caption{\label{fig-vsFDM}
Comparison between results of the FDM and PINN.
Upper: the spatio-temporal variation of the thickness $h(t,x)$ 
predicted by (a) Finite Differences and (b) PINN.
Lower: the selected snapshots of the 
instantaneous spatial thickness distribution at (c) $t=0$, (d) $t=0.5$, and (e) $t=1.0$.
}
\end{figure*}

\begin{figure}[tbp]
\centering\includegraphics[width=0.6\textwidth]{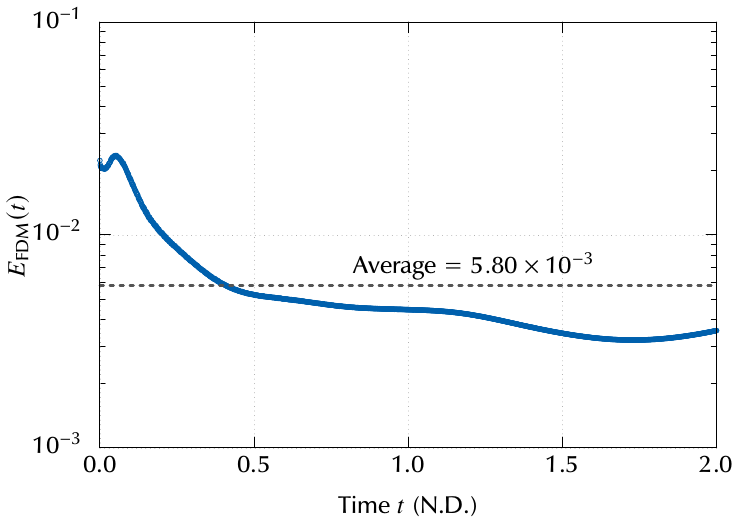}
\caption{\label{fig-RMSRE}
Instantaneous root mean squared relative error $E_\text{FDM}$ between the results of the PINN and FDM.
}
\end{figure}

First, we demonstrate a representative case for the condition shown in \cref{tab-setup1,tab-setup2}.
The blue line in \cref{fig-lossEq124} shows the history of the loss function during training.
The loss function $J$ is reduced
below \num{1e-3} after approximately \num{1e6} iterations.
Because the loss function $J$ is evaluated using only the selected sampling points $(t_i, x_i)$,
it is not guaranteed that the PINN can predict an accurate solution for any given point, 
even if the value of the loss function is sufficiently small.
Thus, the PINN solutions were compared in detail with results obtained using the FDM.
\Cref{fig-vsFDM} summarizes the result of the representative case.
The color contours in \cref{fig-vsFDM}(a,b) show the spatio-temporal solutions $h(t,x)$,
where the color contours are scaled as $\log_{10}(h)$.
For comparison, the results calculated by the FDM are shown in \cref{fig-vsFDM}(a),
and the results predicted by the PINN are in \cref{fig-vsFDM}(b).
The lower three sub-figures \cref{fig-vsFDM}(c,d,e) are
selected snapshots of thickness variations at different time instants $t$ = \numlist{0;0.5;1.0}.
The training of the PINN is sufficiently converged,
and the trained PINN can predict solutions sufficiently close to those of the FDM, 
even for the points $(t, x)$ that are not provided as training data.
In order to discuss the accuracy quantitatively,
the difference between the results of the PINN and FDM is 
evaluated by the following root mean squared relative error:
\begin{equation}
E_\text{FDM}(t) = \sqrt{\frac{1}{N_x}\sum_i^{N_x} \left(\frac{h_\text{PINN}\left(t, x_i\right) - h_\text{FDM}\left(t, x_i\right)}{h_\text{FDM}\left(t, x_i \right)} \right)^2},  \label{eq-RMS}
\end{equation}
where $h_\text{PINN}$ and $h_\text{FDM}$ are the solutions predicted by the PINN and FDM, respectively.
\Cref{fig-RMSRE} shows the instantaneous $E_\text{FDM}$ as a function of time $t$.
The time-averaged value of the $E_\text{FDM}$ was \num{5.80e-3}, as indicated in \cref{fig-RMSRE},
which can be considered sufficiently small.
The maximum value of $E_\text{FDM}$ occurs at $t=0$, even if the initial condition is applied.
In classical numerical methods,
the discretized governing equation is numerically integrated in time, starting from the initial condition.
In the PINN, the initial condition is given as part of the loss function to be minimized;
thus, it is not guaranteed that the initial condition is always satisfied.
In the present problem, in which the initial thickness undulation is planarized by Laplace pressure,
the temporal thickness variation is fastest at $t=0$, 
and this is the reason why the $E_\text{FDM}$ reaches its maximum at $t=0$.

\begin{table}[bp]
\caption{\label{tab-calTime}
Execution environments and elapsed calculation time required for the FDM and PINNs.
}
\centering
\begin{tabular}{lcccc}
\toprule
                       &                       & \multicolumn{3}{c}{PINN} \\
\cline{3-5}
Method                 &   FDM                 & 4th-order       & 2nd-order  & 1st-order  \\
\midrule
CPU/GPU                &   Xeon E5-2603v4      & \multicolumn{3}{c}{NVIDIA A100}          \\
Performance            &   \num{163.2} GFLOPS  & \multicolumn{3}{c}{\num{19.5} TFLOPS}     \\
\hline
Time steps/epochs      &   \num{5e4}           & \num{1e6}           &   \num{1e6}         & \num{1e6}            \\
\multirow{2}{*}{Computation time}
                       &   \SI{25049}{s}       &  \SI{140277}{s}     & \SI{48587}{s}       & \SI{36034}{s}        \\
                       &   (\SI{6.96}{h})      &  (\SI{38.97}{h})    & (\SI{13.50}{h})     & (\SI{10.01}{h})      \\
Time-averaged $E_\text{FDM}$ &   ---           & \num{5.80e-3}       & \num{4.97e-3}       & \num{4.99e-3}        \\ 
\bottomrule
\end{tabular}
\end{table}

\Cref{tab-calTime} shows the computational time required for the time integration of the FDM and training iterations in the PINNs.
Because of the difference in the computational environments for the FDM and PINN,
it is difficult to compare the computational time quantitatively.
The code for the FDM is parallelized with OpenMP and executed in a shared memory CPU environment,
whereas the code for the PINN is executed with the assistance of a GPU.
Nevertheless, we note that training the PINN is computationally expensive.
Once training of the PINN converges,
the solution $h(t,x)$ can be predicted immediately without any time integration.
However, computational time for training is regarded as a major issue from a practical viewpoint.
In the field of machine learning, many computationally efficient methods for training have been developed,
and such methods can be easily applied to the framework of the PINN.
In this study, we consider a PINN-specific method to reduce the computational requirements for training.

\subsection{Effect of order of derivatives in PDEs}
\begin{figure}[tbp]
\centering\includegraphics[width=0.6\textwidth]{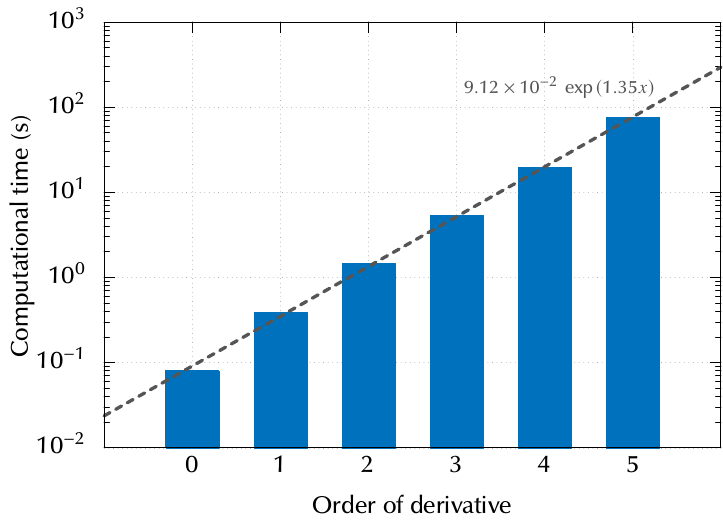}
\caption{\label{fig-timeForAD}
Computational time for automatic differentiation as a function of the order of derivatives.
The number of hidden layers is $N_\ell = 30$, 
the number of neurons in each layer is $N_h = 200$,
and the size of the data set is $N_f = \num{20000}$.
The dashed line depicts the fitted exponential function
$\num{9.12e-2}\exp(\num{1.35}x)$.
}
\end{figure}
Automatic differentiation (AD) can be regarded as a bottleneck in the computational time for the PINN.
In this subsection, we investigate the computational cost for AD, especially the dependence on the order of the derivatives in PDE.
To this end, the computational time for AD is evaluated by the code whose main part is illustrated in the following Python code snippet:
\begin{lstlisting}[language=Python, basicstyle={\ttfamily}, keywordstyle={\bfseries}, columns=fixed]
def auto_diff(x, order):
    h = neural_net(x, weights, biases)
    for i in range(order):
       h = tf.gradients(h, x)[0]
    return h
\end{lstlisting}
This code is executed using \texttt{tensorflow-1.8} by supplying the data set $\{t_i,x_i\}_{i=1}^{N_f}$ of $N_f = \num{20000}$ 
with a network size of ($N_\ell, N_h) = (30, 200)$.
\Cref{fig-timeForAD} shows the computational time required for AD as a function of the order of derivatives.
The dashed line shows the fitted function, which clearly indicates an exponential dependence of computational time on the order of the derivative.
The time required for the 4th-order derivative is approximately 10 times longer than that for the 2nd-order.
As far as the authors are aware,
all the AD implementations available to date calculate the higher-order derivatives by recursion of the single-order AD. 
Therefore, the computational time for AD is exponentially dependent on the order of the derivatives.
Regarding the PINN for the present liquid film flow problem, 
AD for the 4th-order spatial derivative seems to dominate the computational time.
In this light, we consider reducing the order of the derivatives 
by dividing the 4th-order PDE into the coupled 2nd-order or 1st-order PDEs 
by introducing the intermediate variables, as shown in \cref{eq-governingEq2,eq-governingEq4}.
The neural network used in this investigation has multiple outputs, as shown in \cref{fig-structure}(b,c).
The training data set is the same as that used with the 4th-order PDE.

The history of the loss function $J$ during training is indicated by the green line in \cref{fig-lossEq124}.
The trajectory and converged value of $J$ for the 2nd-order and 1st-order PDEs are almost the same as those for the 4th-order PDE.
With respect to the difference between the PINN and the FDM,
the evaluated values of $E_\text{FDM}$ were almost the same as that for the 4th-order PDE.
The entire computational time required for training is summarized in \cref{tab-calTime}.
The computational time taken for the training of the 2nd-order PINN is approximately \SI{35}{\percent} of that for the 4th-order,
whereas for the 1st-order PINN the computational time is approximately \SI{26}{\percent}.

By division of the PDE, 
the maximum order of the derivative decreases, whereas the number of terms to be evaluated increases.
From the present investigation, 
it was confirmed that the decrease of the order of derivatives in the AD 
outweighs the increase in the number of 
equations,
resulting in a net decrease in computational time required.
This competition is strongly dependent on the efficiency of the algorithm for the higher-order AD, which may be improved in future.

\subsection{Precision of floating-point numbers}
\begin{figure}[tbp]
\centering\includegraphics[width=0.7\textwidth]{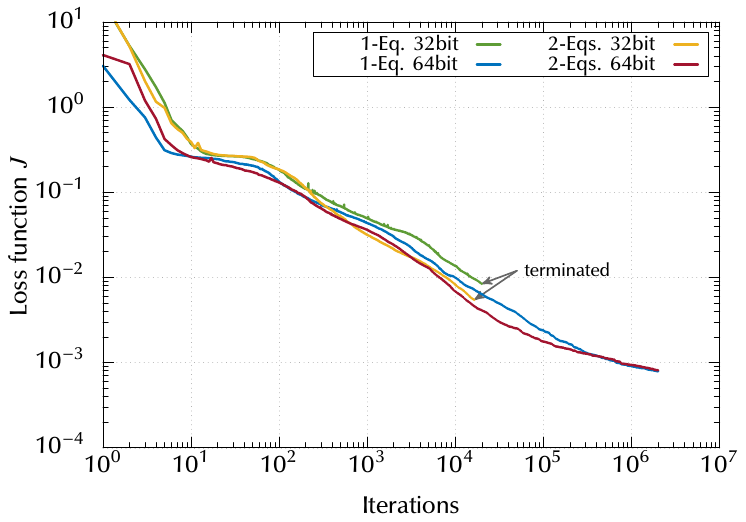}
\caption{\label{fig-loss_FP32_64}
History of the loss function $J$ during training for 
different floating-point precision 
and different orders of derivatives.
}
\end{figure}
In this section, we show how the precision of floating-point operations affects the training performance of the PINN.
For the 1-dimensional Burgers equation, which is a 2nd-order PDE,
Raissi \etal showed that a PINN can be trained using single-precision floating-point operations (FP32) with an activation function of $\tanh$,
and can achieve sufficient accuracy compared with the analytical solution \cite{Raissi_JCP_2019}.
For the present study, PINN training failed when FP32 precision was selected.

\Cref{fig-loss_FP32_64} indicates the history of the loss function $J$ during training
for the cases with different floating-point precision (FP32/FP64) and different orders of derivatives (2nd-/4th- order).
The loss functions $J$ for all cases show similar behavior,
allowing for differences in the initial values because of the randomly initialized weights.
Computation for the two FP64 cases continued until the number of iterations reached the selected maximum value of \num{2e6}.
The FP32 cases, on the other hand, were terminated after approximately \num{2e4} iterations;
nevertheless, the loss function $J$ can potentially be reduced.
It should be noted that
the FP32 computation failed even when the governing equation was divided into two 2nd-order PDEs, 
which are denoted as the `2-Eqs. 32bit' case in \cref{fig-loss_FP32_64}.
In the following, we consider the reason why the FP32 computation fails.

In the present work, 
the L-BFGS-B method in the \texttt{SciPy} implementation is employed for optimization,
and this algorithm stops its iteration when either or both of the following conditions are satisfied:
\begin{subequations}
\begin{gather}
\frac{J^k - J^{k+1}}{\max \left\{ \left| J^k \right|, \left| J^{k+1} \right|, 1  \right\}} \leqslant \texttt{ftol}, \label{eq-stopF}  \\
\max_i{ \left\{ \left| g_i \right| \right\} } \leqslant \texttt{gtol},  \label{eq-stopG}
\end{gather}
\end{subequations}
where $J$ is the loss function,  $k$ is an index for iterations, and $g_i$ is the i-th component of the projected gradient of $J$.
\texttt{ftol} and \texttt{gtol} are the tolerance values,
selected as \texttt{ftol} = \num{2.2e-16} and \texttt{gtol} = \num{1e-15}, respectively.
In the present FP32 computations, the iteration is stopped by the condition \cref{eq-stopF};
nevertheless, the projected gradient $g$ is still much larger than the tolerance \texttt{gtol}.
In general, the gradient-based optimization method 
first finds a descent direction $\bol{p}_k$ along which the loss function $J$ will be reduced,
and then computes a step size $\alpha_k$ that adequately reduces $J(\bol{x}_k + \alpha_k\bol{p}_k)$ relative to $J(\bol{x})$.
The step size $\alpha_k$ is computed using line search methods, which have several variations.
In an inexact line search, which is implemented in \texttt{SciPy},
the step size $\alpha_k$ is determined such that the sufficient decrease condition and the curvature condition are satisfied \cite{Nocedal-book-2006}.
In the present PINN computations with FP32,
the step size $\alpha_k$ becomes as small as $\order{10^{-10}}$ at termination,
thus the update of the loss function $J$ becomes vanishingly small
and the stopping condition \cref{eq-stopF} is satisfied.
Such a small value of $\alpha_k$ might occur when the loss function $J$ has nearly stationary points and the parameter $\bol{x}$ gets trapped therein.
The FP32 computation is likely to lose precision in the evaluation of $J$ 
due to the cancellation of significant digits.

It is well known that the activation function \texttt{tanh} is likely to suffer from the so-called vanishing gradient problem \cite{Hochreiter1998}.
The vanishing gradient problem occurs
because the \texttt{tanh} function has gradients in a narrow range, and backpropagation computes gradients using the chain rule.
Thus, the gradient decreases exponentially during backpropagation, and the weights in the early layers receive vanishingly small updates.
The PINN for the present work may suffer from the vanishing gradient problem,
because the number of chain rule operations
increases exponentially as the order of the derivative increases, as shown in \cref{fig-timeForAD}.

The use of rectifier functions such as \texttt{ReLU}
is expected to reduce the effect of the vanishing gradient problem, because these functions saturate in only one direction.
To investigate the effect of the activation function,
we trained the PINN for the 4th-order PDE of \cref{eq-governing} with FP32 precision,
by changing the activation function 
\texttt{tanh} to the \texttt{swish} function, which is defined as
\begin{equation}
a(z) = \frac{z}{1+\exp(-z)}.
\end{equation}
However, training with FP32 using this activation function failed for the same reason as for \texttt{tanh}.
Although there are many other rectifier functions,
their effects have not been investigated in the present study.
Even if some activation function enables successful training with FP32 precision,
there still exists problem of selecting an appropriate activation function.

As described above, the FP32 computation failed 
even when the governing equation is divided into two 2nd-order PDEs.
This suggests that 
the failure of FP32 computation is not caused by higher-order automatic differentiation,
which requires many multiplications in the chain rule.

Based on the investigations described so far, 
the failure of PINN training with FP32 precision seems to be caused by 
the loss of significant digits during the line search  
when the loss function has nearly stationary points.

The failure of training with FP32 precision can potentially be avoided 
by the selection of an appropriate optimization method or the adjustment of the numerical parameters for optimization.
However, from a comprehensive standpoint, 
FP32 computation has a risk of failure for the present problem compared with FP64.

\subsection{Density distribution of collocation points}\label{sec-collocation}
\begin{figure}[tbp]
\centering\includegraphics[width=0.7\textwidth]{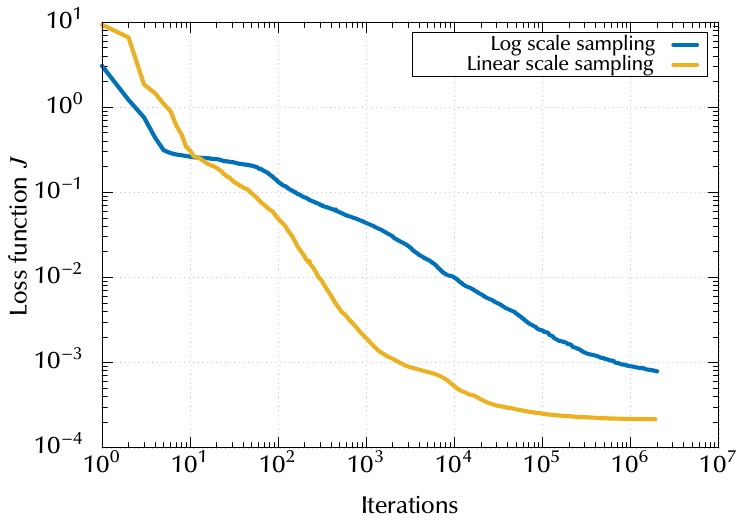}
\caption{\label{fig-lossLinear}
History of the loss function $J$ during training 
for different sampling density of the collocation points.
The blue line indicates the result of log-scale time sampling,
whereas the orange line represents linear-scale time sampling.
}
\end{figure}
\begin{figure}[tbp]
\centering\includegraphics[width=0.7\textwidth]{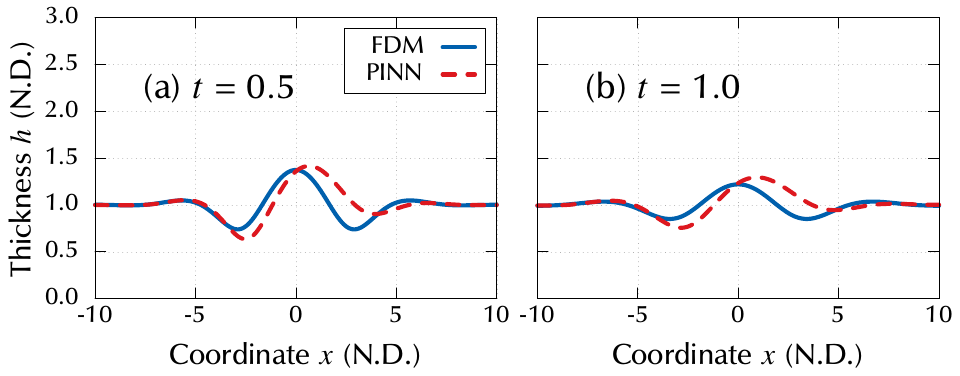}
\caption{\label{fig-linearSnapshots}
Selected snapshots of the 
instantaneous spatial thickness distribution at (a) $t=0.5$ and (b) $t=1.0$,
for the case in which collocation points are sampled on the linear scale.
}
\end{figure}
It was found that
the convergence and accuracy of the PINN are strongly affected by the sampling density of the collocation points.
We considered two types of scaling of points in time, as described in \cref{sec-trainingData}.
One of them is a linear scale sampling, where the time points are evenly spaced as in \cref{eq-linScale}.
The other is a log scale sampling, where the time points are evenly spaced on a logarithmic scale as in \cref{eq-logScale}.
The examples of the collocation points are shown in \cref{fig-samplePoints}.
The convergence and the accuracy of the PINN are investigated for both sampling types.
\Cref{fig-lossLinear} indicates the history of the loss function $J$ during the training.
The blue line corresponds to log scale sampling, which is the same as the green line in \cref{fig-lossEq124}.
The orange line in \cref{fig-lossLinear} is the result of linear scale sampling.
It looks at first glance as though the linear scale provides
faster convergence and a smaller final value of $J$ compared to the log scale.
However, it should be noted that
the solutions predicted by the trained PINN are not guaranteed to be accurate for any given point $(t,x)$
even if the loss function $J$ converged to a sufficiently small value.
Because the loss function $J$ is evaluated using only the selected sampling points of $(t_i, x_i)$,
the sampling of the collocation points significantly affects the absolute value of $J$.
Thus, the solutions are compared with the result of the FDM.
\Cref{fig-linearSnapshots} shows the selected snapshots of the thickness distributions 
predicted by the trained PINN for the case of linear scale sampling.
Although the loss function $J$ converges and the final value of $J$ for the case of the linear scale is smaller than that for the log scale, 
the predicted solutions differ unacceptably from the result of the FDM.

Appropriate sampling of collocation points surely depends on the problem to be solved.
It seems that the collocation points should be dense 
where the solution has large variations in spatial and/or temporal directions.
For the present problem, log-scale sampling is effective
because the rapid thickness variation occurs at earlier time ranges.

Training efficiency might be further improved
by more optimized sampling of the collocation points.
One possible approach is a self-adaptive approach, which has been proposed by McClenny \& Braga-Neto \cite{McClenny2020}.
In their approach, the trainable weights are introduced as a soft multiplicative mask for the mean squared error for each collocation point, 
and these weights are optimized concurrently with the network weights.

\subsection{Effect of network size}
\begin{figure}[tbp]
\centering\includegraphics[width=0.7\textwidth]{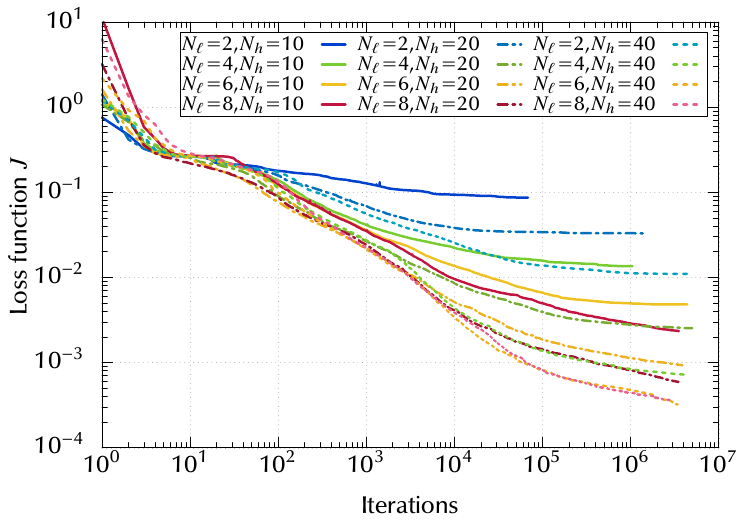}
\caption{\label{fig-loss_paramSurvey}
History of the loss function $J$ during training
for different 
numbers of hidden layers $N_\ell$ and
numbers of neurons in each hidden layer $N_h$.
}
\end{figure}
\begin{table}[tbp]
\caption{\label{tab-finalLossParamSurvey}
Final values of the loss function $J$ when the optimization iteration is terminated.
}
\centering
\begin{tabular}{ccccc}
\toprule
         &    &  \multicolumn{3}{c}{$N_h$} \\
\cline{3-5} 
         &    &  10 & 20 & 40  \\
\midrule
\multirow{4}{*}{$N_\ell$}
    & 2 & 
 \num[round-mode=figures,round-precision=3]{8.662902E-02} &
 \num[round-mode=figures,round-precision=3]{3.306703E-02} &
 \num[round-mode=figures,round-precision=3]{1.103415E-02} \\
    & 4 & 
 \num[round-mode=figures,round-precision=3]{1.361881E-02} &
 \num[round-mode=figures,round-precision=3]{2.538867E-03} &
 \num[round-mode=figures,round-precision=3]{7.182849E-04} \\
    & 6 & 
 \num[round-mode=figures,round-precision=3]{4.825130E-03} &
 \num[round-mode=figures,round-precision=3]{9.341770E-04} &
 \num[round-mode=figures,round-precision=3]{3.224258E-04} \\
    & 8 & 
 \num[round-mode=figures,round-precision=3]{2.349463E-03} &
 \num[round-mode=figures,round-precision=3]{6.364839E-04} &
 \num[round-mode=figures,round-precision=3]{3.649724E-04} \\
\bottomrule
\end{tabular}
\end{table}
\begin{figure}[tbp]
\centering\includegraphics[width=0.6\textwidth]{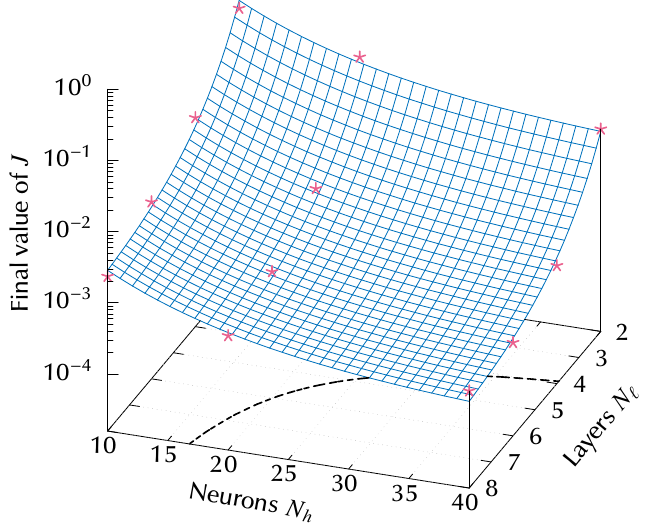}
\caption{\label{fig-finalLoss_paramSurvey_splot}
Final values of the loss function $J$ 
when the optimization iteration is terminated,
as a function of the number of neurons in each hidden layer $N_h$
and the number of hidden layers $N_\ell$.
The red stars are calculated cases, 
and the blue surface indicates a fitted function
$ \log_{10} \left[ J_\text{fit}(N_\ell, N_h) \right]
 = a_0 + a_1 \exp(-N_\ell / {b_1}) + a_2 \exp(-N_h / {b_2})$
with coefficients
$a_0 = \num[round-mode=figures,round-precision=3]{-3.80809}$,
$a_1 = \num[round-mode=figures,round-precision=3]{ 4.39276}$,
$b_1 = \num[round-mode=figures,round-precision=3]{ 2.09156}$,
$a_2 = \num[round-mode=figures,round-precision=3]{ 2.43163}$, and
$b_2 = \num[round-mode=figures,round-precision=3]{ 13.6543}$.
The final sum of squares of residuals is
$R^2 = \num[round-mode=figures,round-precision=3]{8.87797e-2}$.
The black dashed line drawn on the base plane indicates
the values of $N_\ell$ and $N_h$ where the loss function reaches $J = 10^{-3}$.
}
\end{figure}

In the results described so far,
the network architecture is kept fixed unless otherwise noted,
as shown in \cref{fig-structure}(a).
In the present study,
the network is composed of fully-connected hidden layers,
thus the network size is simply determined by
the number of the hidden layers $N_\ell$
and the number of neurons $N_h$ in each layer.
With a network size of $N_\ell = 8$ and $N_h = 20$,
it has been shown that the PINN can be trained with sufficient accuracy.

The PINN can be regarded as a
kind of fitting of the nonlinear mapping between inputs (time and space) and outputs (solution)
by the linear combinations and simple nonlinear activation functions.
Therefore, how complicatedly the mapping can be reproduced must depend on
the number of degrees of freedom of the neural network.
In this section, the effect of the network size is investigated by varying $N_\ell$ and $N_h$.

\cref{fig-loss_paramSurvey} shows the history of the loss function $J$ during training.
The final values of $J$ after training is terminated are summarized in \cref{tab-finalLossParamSurvey}.
The trajectory and converged values of $J$ are strongly dependent on the network sizes $N_\ell$ and $N_h$.
The definition of the loss function $J$ itself is not dependent on the network size,
and the number of sampling points is kept fixed for all cases
as $N_f = 6000$, $N_b = 1000$ and $N_o = 400$.
Thus, the difference in the value of $J$ can be regarded as an effect of the network size.
As can be easily predicted,
for larger values of $N_\ell$ and $N_h$,
the loss functions $J$ converged to smaller values.

As described in \cref{sec-repCase},
the PINN with network size $(N_\ell,N_h)$ = $(8,20)$ can be trained successfully.
The root mean square relative error 
between the PINN prediction and the result of the FDM was \num{5.80e-3} on average.
With this network size, the final value of $J$ was \num{6.36e-4},
which can be considered as sufficient convergence.
Among all variations of $N_\ell$ and $N_h$ in the present work,
the final values of the loss function surpassed $J \leqslant 10^{-3}$
for five cases of  
$(N_\ell,N_h)$ = $(6,20)$, $(8,20)$, $(4,40)$, $(6,40)$, and $(8,40)$.

To estimate the necessary values
for $N_\ell$ and $N_h$
such that sufficient convergence of $J$ can be achieved,
the calculated data are fitted by the following function
\begin{equation}
\log_{10} \left[ 
 J_\text{fit} \left(N_\ell, N_h \right) \right]
= a_0 + a_1 \exp\left(-\frac{N_\ell}{b_1} \right) + a_2 \exp \left(-\frac{N_h}{b_2}\right), \label{eq-fit}
\end{equation}
where the coefficients are determined by the least square method as
$a_0 = \num[round-mode=figures,round-precision=3]{-3.80809}$,
$a_1 = \num[round-mode=figures,round-precision=3]{ 4.39276}$,
$b_1 = \num[round-mode=figures,round-precision=3]{ 2.09156}$,
$a_2 = \num[round-mode=figures,round-precision=3]{ 2.43163}$, and
$b_2 = \num[round-mode=figures,round-precision=3]{ 13.6543}$,
with the residual sum of squares of $R^2 = \num[round-mode=figures,round-precision=3]{8.87797e-2}$.
In \cref{fig-finalLoss_paramSurvey_splot},
the final values of the loss function $J$
are plotted three-dimensionally by the red stars.
In addition, the fitted function \cref{eq-fit} is represented by the blue surface.
From the fitting function \cref{eq-fit},
we can estimate the necessary values for $N_\ell$ and $N_h$
so that sufficient convergence of $J$ can be achieved.
This necessary condition is plotted in \cref{fig-finalLoss_paramSurvey_splot} as a black dashed line,
by considering sufficient convergence to be $J \leqslant 10^{-3}$.
At the bounds of the range investigated in the present study,
this condition corresponds to
$N_\ell = \num{3.9}$ for $N_h = 40$
and
$N_h = \num{16.7}$ for $N_\ell = 8$.
When the degrees of freedom in the network exceed the above condition,
the PINN can reproduce
the nonlinear mapping between $x,t$, and $h$ sufficiently with the present governing equation.


\subsection{Robustness of trained model}
When the physics-informed neural network is applied to the practical problems,
for instance, the data assimilation, the input data may include noise.
For such a situation, 
the solutions predicted by the trained PINN should not largely deviate from those for the clean (noise-free) inputs.
For the trained PINN of the present work,
the effect of the noise was investigated as follows.

\begin{figure}[tbp]
\centering\includegraphics[width=0.6\textwidth]{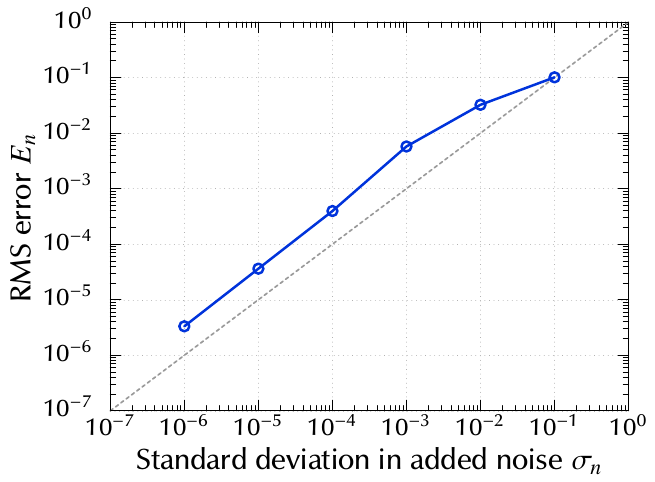}
\caption{\label{fig-noisyInput}
Root mean squared error $E_n$ defined by \cref{eq-En} as a function of the standard deviation of noise added to the inputs.
The dashed line indicates the $y=x$.
}
\end{figure}

First, the evaluation points $x_i, t_i$ are generated with equally-spaced grids of $(N_x, N_t) = (1000,400)$.
The noisy inputs are generated by adding the normally-distributed noises:
\begin{equation}
\delta x, \delta t   \sim N\left(0, \sigma_n \right), 
\end{equation}
where $\delta x$ and $\delta t$ are the noise for $x$ and $t$, respectively.
$\sigma_n$ is the standard deviation of the normal distribution.
The solutions of the PINN are calculated for both of the noisy and clean inputs,
then the difference of them $\Delta h_i$ and the root mean squared error $E_n$ are defined as
\begin{align}
\Delta h_i  &= h_\text{PINN}\left(x_i + \delta x_i, t_i  + \delta t_i  \right) - h_\text{PINN}\left(x_i, t_i \right), \\
E_n         &= \left[ \frac{1}{N_e} \sum_{i=1}^{N_e} \left( \Delta h_i  \right)^2 \right]^{\frac{1}{2}},   \label{eq-En}
\end{align}
where $N_e = N_x \times N_t$ is the total number of the evaluation points.
The values of $E_n$ are calculated by changing the noise corruption levels.
\Cref{fig-noisyInput} indicates the 
root mean squared error $E_n$ as a function of the standard deviation $\sigma_n$ for the applied noise.
It can be seen that the $E_n$ is approximately the same order of magnitude with the $\sigma_n$.
This means that the thickness for the noise-corrupted inputs can be obtained
with the error of the the same order of magnitude of the noise included.
It can be regarded that the PINN of the present work was not overfitted for the training dataset,
and it can provide the accurate solutions even for the inputs that was not encountered during the training.

\begin{figure}[tbp]
\centering\includegraphics[width=0.8\textwidth]{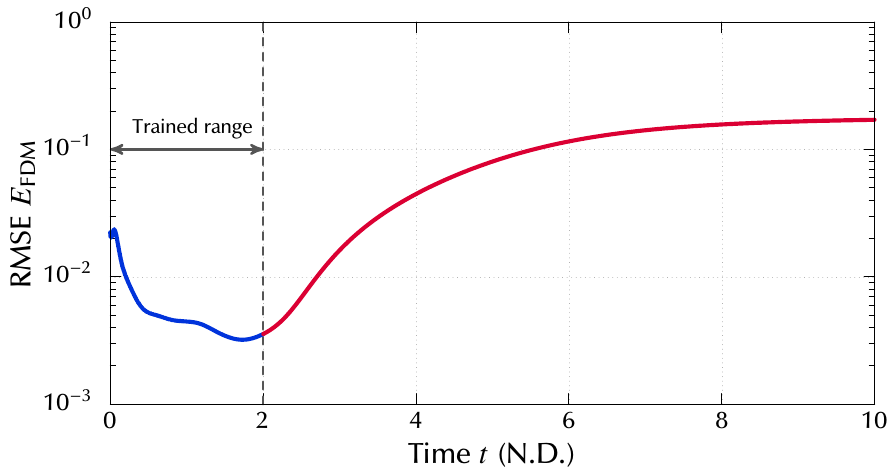}
\caption{\label{fig-relativeOut}
Instantaneous root mean squared relative error $E_\text{FDM}$ between the results of the PINN and FDM.
The collocation time points $t_i$ for the training are sampled from the time horizon $0\leqslant t \leqslant 2$, which is indicated by the blue line.
}
\vskip1em
\centering\includegraphics[width=0.8\textwidth]{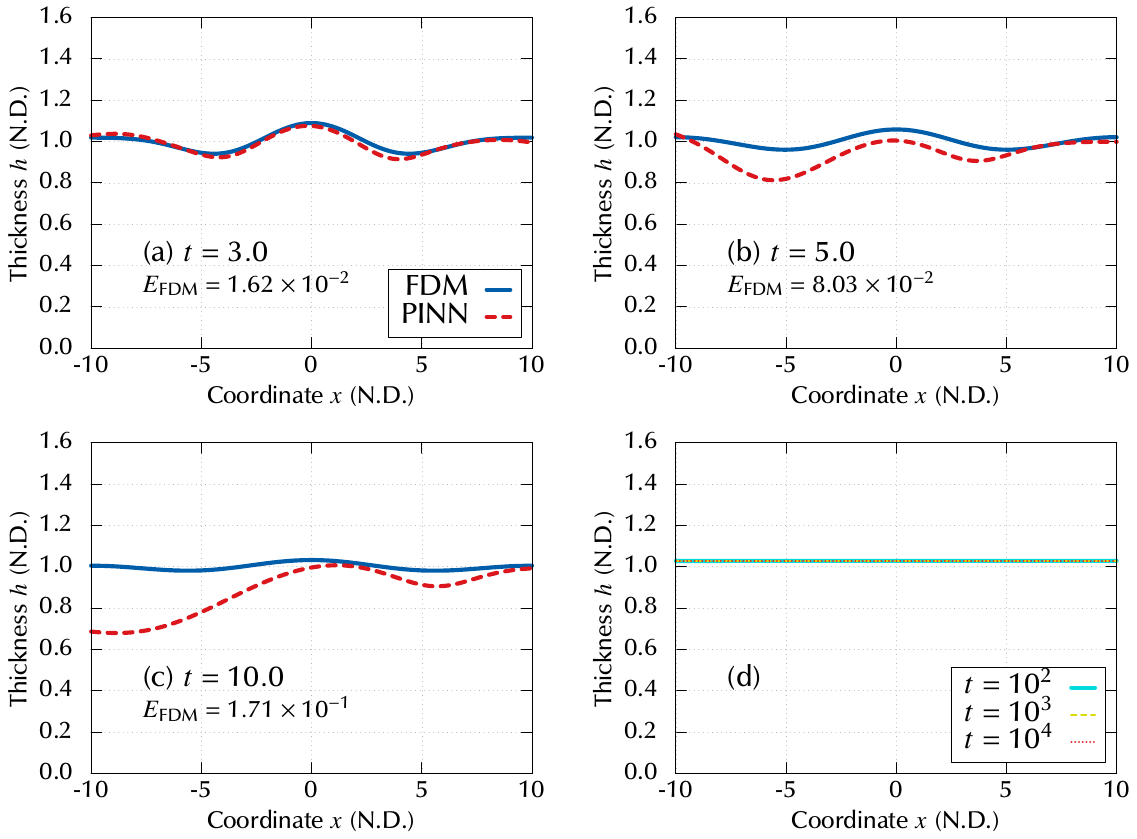}
\caption{\label{fig-outRange}
Selected snapshots of the instantaneous spatial thickness distribution 
for the time instances outside of the time horizon of collocation points for the training.
(a) $t = 3.0$,
(b) $t = 5.0$, 
(c) $t =10.0$, and
(d) $t= 10^2, 10^3, 10^4$.
In the sub-figures except (d), 
the solutions obtained by the finite difference method are also plotted for the comparison.
}
\end{figure}
The robustness of the trained PINN was also investigated from another viewpoint:
how the PINN predicts the solution when the time input is outside of the training data regime.
In general function fitting methods,
the function is fitted for the values that span a specified dataset only.
For the outside of the dataset, output of the fitted function are not guaranteed,
and the output values might be largely different from those expected from the original dataset.
Although the PINN is one of the function fittings, 
in contrast to the conventional methods, 
the loss function to be minimized is based not on the function value itself but on the governing equation.
The weights of the PINN are optimized so that 
not only the field variable, 
but also the time and spatial derivatives satisfy the governing equation.
Thus the solutions of the PINN may satisfy the governing equation in certain degree, even for the time outside of the dataset.

In this light, the prediction of the PINN were evaluated for the wide time range.
\Cref{fig-relativeOut} shows the instantaneous root mean squared relative error $E_\text{FDM}$
between the solutions of the PINN and the FDM, which is defined as \cref{eq-RMS}.
The collocation time points $t_i$ for the training are sampled from the time horizon $0\leqslant t \leqslant 2$, which is indicated by the blue line in \cref{fig-relativeOut}.
The plot for this time range is the same as that indicated in \cref{fig-RMSRE}.
The red line in \cref{fig-relativeOut} indicates the error $E_\text{FDM}$
for the time range $2 < t$, which is outside of the training dataset.
Although the value of $E_\text{FDM}$ increases with time $t$,
it does not diverge and takes at most $E_\text{FDM} = \num{1.71e-1}$ in the evaluated time range $t \leqslant 10$.
\Cref{fig-outRange} shows the selected snapshots of the instantaneous spatial thickness distribution $h$,
where the solutions obtained by the FDM are also plotted.
It can be seen that 
the solution of the PINN at $t=3.0$ is close to that obtained by the FDM (\cref{fig-outRange}(a)).
For the later time range (\cref{fig-outRange}(b,c)),
the solutions of the PINN clearly differ from those of the FDM.
Though in the qualitative sense,
some similarities can be recognized between solutions of the PINN and the FDM:
the extrema and their locations, gradient at the boundaries.
The spatial symmetry in $h$ is not retained for the later time range.

As shown in \cref{fig-outRange}(d),
the solutions of the PINN are also evaluated for the very large time value.
For the sinusoidal deformation of the wavelength $\lambda_m$, 
the time constant for the decay by the Laplace pressure can be derived as
\begin{equation}
\tau =\frac{3\mu}{\sigma h_0^3}\frac{\lambda_m^4}{16\pi^4}, \label{eq-tau}
\end{equation}
from the analysis of the linearized governing equation.
Using the time scale $t_\ast = 3\mu \lambda^4 / \sigma h_0^3$ of the current problem, as described in \cref{tab-scales},
the nondimensional form of \cref{eq-tau} can be written as $\hat{\tau} = \hat{\lambda}_m^4 / 16\pi^4$, where $\hat{\lambda}_m$ stands for the nondimensional wavelength.
In this problem, the longest wavelength can be estimated as $\hat{\lambda}_m \approx 2 x_L = 20$,
which is corresponding to the time constant $\hat{\tau} \approx 100$.
Therefore, the time instances are selected as $t=10^2$, $10^3$ and $10^4$.
For these time instances, the thickness must be flattened by the Laplace pressure.
Accordingly, this situation can be confirmed in \cref{fig-outRange}(d), 
where the solutions of the PINN are completely flat distribution.
It is noted that the three plots for $t=10^2, 10^3, 10^4$ are overlapped in the figure.
By integrating the initial condition \cref{eq-Ini},
the exact value of the averaged thickness can be analytically calculated as $h_\text{ave} = 1$.
The flat thickness at $t=10^4$ is $h = \num{1.028}$ (\cref{fig-outRange}(d)),
which is \SI{2.8}{\percent} higher than the exact solution.
%

\section{Concluding remarks}
In the present study, 
a physics-informed neural network  was applied to a partial differential equation of liquid film flows.
The PDE considered is the time evolution of the thickness distribution $h(x,t)$ owing to Laplace pressure that
involves the 4th-order spatial derivative and 4th-order nonlinear term.
Even for such a PDE, it was confirmed that the PINN can predict the solutions with sufficient accuracy.
Nevertheless, some aspects are needed to improve training convergence and accuracy of the solutions.

Calculation of the automatic differentiation (AD)
dominates the computational time required for training, 
and becomes exponentially longer as the order of derivatives increases.
By splitting the original 4th-order single PDE into lower-order coupled PDEs,
the computational time for a single training iteration was greatly reduced.
The precision of the floating-point numbers is a critical issue for the present PDE.
When the calculation is executed with FP32 (single precision), 
training terminated due to the loss of significant digits,
despite the loss functions not being reduced sufficiently.
The density distribution of the collocation points for the training data also significantly affected training convergence.
For the problem considered in the present study,
convergence was improved by allowing the sampling density to be higher in earlier time ranges, 
where the rapid diffusion of the thickness occurs.
To clarify the degrees of freedom in the neural network
required to reproduce the nonlinear mapping between input and output,
the accuracy of the PINN prediction was
investigated by varying the number of hidden layers $N_\ell$ and the number of neurons in each layer $N_h$.
From the fitting function \cref{eq-fit}, the necessary values for $N_\ell$ and $N_h$ are evaluated.

From a comprehensive point of view, 
the original form of the PINN requires a lot of attention to ensure successful training.
This is especially true in the case of the higher-order PDEs dealt with the present study.
A lot of preliminary investigation is needed to find appropriate configuration
for the network size and selection of the collocation points.
In addition, the long computational time required for training may also be practical bottleneck.
Further research is needed to overcome these problems.
Regarding computational time,
fast algorithms and an efficient implementations for automatic differentiation, especially for higher orders of derivative, are needed.
Regarding the selection of the collocation points,
it is expected that self-adaptive approaches will improve the training performance.
Another possible approach is 
an evolutional deep neural network (EDNN), which has been recently proposed by Du \& Zaki \cite{Du2021}.
In EDNN, the neural network is trained only for the initial condition,
and the network weights are updated based on the time evolutional equation derived from the governing equation.
Use of this improvement is expected to contribute to practical application of the PINN.

\section*{Acknowledgments}
This work was supported by JSPS KAKENHI (Grant Number JP19K04175).
The calculations shown in the present work were executed on 
the SGI Rackable C2112-4GP3/C1102-GP8 (Reedbush-U/H) 
and
the FUJITSU Supercomputer PRIMEHPC FX1000 and FUJITSU Server PRIMERGY GX2570 (Wisteira/BDEC-01)
in the Information Technology Center, The University of Tokyo.
The computational resource of the HPE SGI 8600 was supported by in the Institute of Statistical Mathematics.



\end{document}